\documentstyle[12pt,pra,aps]{revtex}       

\title{On quantum logic operations based on 
photon-exchange interactions in an ensemble of
non-interacting atoms}
  
\author{M.~Fleischhauer}
\address{ITAMP,
Harvard-Smithsonian Center for Astrophysics, Cambridge, MA 02138}

\date{\today}
   
\begin{document}

\maketitle
 \tighten
\begin{abstract}
The recently proposed idea to generate entanglement between photon states
via exchange interactions  in an ensemble of atoms
(J.~D.~Franson and T.~B.~Pitman, Phys. Rev. A {\bf 60}, 917 (1999) and
J.~D.~Franson {\it et al.}, (quant-ph/9912121)) is discussed
using an $S$-matix approach.
It is shown that if the nonlinear response of the atoms is negligible
and no  additional atom--atom interactions are present, exchange interactions
cannot produce entanglement between photons states in a process that
returns the atoms to their initial state. 
Entanglement generation requires the presence of a nonlinear atomic
 response or atom--atom interactions. 
\end{abstract}



\section{Introduction}


In some recent papers Franson {\it et al.}, \cite{Fr1,Fr2} suggested  
that exchange interactions of two photons in a macroscopic ensemble
of identical, non-interacting atoms could lead to large conditional 
phase shifts. 
In contrast to ``conventional'' nonlinear optics 
which requires scattering of both photons from the same atom,
exchange interactions are present even when the two photons interact with
different atoms. This makes them much more likely to occur in a dense medium.
The large magnitude of the predicted conditional phase shifts
would make such systems very attractive for quantum logical operation.
However, whether or not exchange interactions are capable of generating 
entanglement between photons has  been subject of some debate
\cite{Kur,Fr3}. In view of the claimed potential advantages,
the requirements and limitations
of the proposed schemes need to be examined. 

In the present note
I want to discuss a special type of 
exchange interactions. In particular I will analyze the possibility
to entangle photon states through interactions in an ensemble
of atoms under the conditions considered in \cite{Fr2}. Namely:
({\it i})  All processes are unitary, i.e. losses are negligible;
 ({\it ii}) The atomic system  
returns to the same state as before the interaction;
({\it iii})
The ``conventional'' nonlinear response of the atoms is
assumed to be negligible;
({\it iv}) It is
assumed that there are no atom--atom interactions, except those through the
quantized radiation modes under consideration. 
Conditions ({\it i}) and ({\it ii}) enshure that the pair of qubits,
represented by the photons
undergoes an effective unitary evolution and is asymptotically
disentangled from the atoms and the environment. 
It will be shown in the following that in a system that fulfills 
conditions ({\it i-iv})
entanglement between a pair of photons
in distinguishable modes can {\it not} be generated. 
Any initially factorizable state
will evolve into a factorizable state.


\section{Model and effective time-evolution operator}


Let me consider the interaction of the quantized radiation field 
with a large number of identical atoms in dipole and rotating-wave
approximation as proposed in \cite{Fr1,Fr2}. In addition to the photon field,
the atoms may be coupled to some external {\it classical} fields to allow for
manipulations of the states after or during the interaction with
the photons.
The Hamiltonian of the system has the following general form
\begin{equation}
H=H_{\rm field}+H_{\rm atom}(t)+V,
\end{equation}
where $H_{\rm field}$ is the free Hamiltonian of the quantized photon field
and $H_{\rm atom}(t)$ is the free Hamiltonian of the atoms including the
interaction with the (time-dependent) external, classical fields. 
For simplicity it is 
assumed that each mode of the photon field couples only to
one atomic transition. It is however straight forward to lift this restriction.
The interaction operator has thus the
following general structure
\begin{eqnarray}
V=- \hbar\sum_k \, g_k\sum_{j=1}^N\left[
{\hat \sigma}_{j,k}^\dagger {\hat a}_k\, f_k({\vec r}_j)
+ {\hat \sigma}_{j,k} {\hat a}_k^\dagger f_k^*({\vec r}_j)\right].
\end{eqnarray}
Here ${\hat a}_k$ and ${\hat a}_k^\dagger$ are annihilation and creation
operators of the photon field. $k$ is a mode index and 
$f_k({\vec r})$ is the associated mode function. $f_k$  
is not restricted to plane waves but could also represent e.g. 
localized wave packets, distinguishable by their arrival time.
 The modes are assumed to be orthogonal,
such that $[a_k,a_{k'}^\dagger]=\delta_{kk'}$.
${\hat \sigma}_{j,k}$ denotes a flip operator of atom $j$ corresponding
to the transition coupled to the mode $k$ with coupling strength $g_k$. 
(Introducing flip operators for different $k$-values
takes into account that the individual modes 
of the quantized field may be coupled to different
dipole transitions.)

It is assumed that initially ($t=t_0$) all atoms are in
their ground states, i.e. the total initial state vector 
has the form
\begin{equation}
|\psi(t_0)\rangle =|\phi(t_0)\rangle\, |g\rangle,
\end{equation}
where $|\phi(t_0)\rangle$ is the initial field state
and $|g\rangle$  the 
collective ground state of the atoms.

The Schr\"odinger-equation 
for the state vector in the interaction picture 
can formally be solved by
\begin{eqnarray}
|\psi(t)\rangle &=& {\sf T}\exp\left\{ -\frac{i}{\hbar}
\int_{t_0}^t\!\!{\rm d}t'\, V(t')\right\} 
|\psi(t_0)\rangle,
\end{eqnarray}
where ${\sf T}$ is the time ordering operator.

It is clear that photon-atom interactions in general entangle
both sub-systems. 
This is however not of interest here. 
The question I want to address is, whether the 
interaction can generate an entangled 
state of the {\sl photons} given that the atomic system returns to its
initial ground state at some time $t_1$. 
Thus we require 
\begin{equation}
|\psi(t_1)\rangle
\longrightarrow|\phi(t_1)\rangle |g\rangle.
\end{equation}
In this case the atomic and photonic components of $|\psi(t_1)\rangle$
factorize and the photonic part is given by
\begin{eqnarray}
|\phi(t_1)\rangle =
\langle g| \, {\sf T}\exp\left\{ -\frac{i}{\hbar}
\int_{t_0}^{t_1}\!\!{\rm d}t'\, V(t')\right\}  |g\rangle
\, |\phi(t_0)\rangle={\sf S}(t_1,t_0)\, |\phi(t_0)\rangle\label{phi}.
\end{eqnarray}
The operator $S$ describes the conditional evolution of the photon field 
when the atomic system returns to its ground state. 

In order to calculate the action of $S$, we
make use of a generalization of the cumulant generation function 
for a classical statistical variable $X$ 
\begin{eqnarray}
\Bigl\langle\exp\bigl\{sX\bigr\}\Bigr\rangle_X =\exp\left\{\sum_{m=0}^\infty
\frac{s^m}{m!}\bigl\langle\bigl\langle X^m\bigr\rangle\bigr\rangle\right\}.
\label{gen}
\end{eqnarray}
Here $\langle\langle X^m\rangle\rangle$ denotes the $m$th order cumulant, i.e.
$\langle\langle X\rangle\rangle =\langle X\rangle$, $\langle\langle X Y
\rangle\rangle =\langle XY\rangle-\langle X\rangle\, \langle Y\rangle$ etc. 
Applying eq.(\ref{gen}) to  ${\sf S}$
yields 
\begin{eqnarray}
{\sf S}(t_1,t_0) &=&
{\sf T}\exp\Biggl\{\int\!\!{\rm d}1\!\int\!\!{\rm d}2\,\, 
{\hat a}_{k_1}^\dagger(\tau_1)\, {\cal P}(1;2)\, {\hat a}_{k_2}(\tau_2)
\nonumber\\
&&\qquad\quad +\int\!\!{\rm d}1\!\int\!\!{\rm d}2\!
\int\!\!{\rm d}3\!\int\!\!{\rm d}4\,\, 
{\hat a}_{k_1}^\dagger(\tau_1){\hat a}_{k_2}^\dagger(\tau_2)
\, {\cal P}^{(2)}(1,2;3,4)\,  {\hat a}_{k_3}(\tau_3)
 {\hat a}_{k_4}(\tau_4) +\cdots\Biggr\}\label{S}
\end{eqnarray}
where  $\int {\rm d}1$ stands for integration over time $\tau_1$ and summation
over the mode index $k_1$. It was assumed here for simplicity that the average
dipole moment of the atoms vanishes.  
\begin{eqnarray}
{\cal P}(1,2) =\sum_j {\cal P}^j(1,2),
\end{eqnarray}
where
\begin{eqnarray}
 {\cal P}^j(1,2)=-
g_k^2 f_{k_1}^*({\vec r}_j) f_{k_2}({\vec r}_j)\,
\Bigl\langle\Bigl\langle {\sf T}{\hat \sigma}_{jk_1}^\dagger(\tau_1) 
{\hat \sigma}_{jk_2}(\tau_2)\Bigr\rangle\Bigr\rangle
\end{eqnarray}
describes  the {\it linear} response of the $j$th atom to
the quantized radiation field. 
The higher-order terms ${\cal P}^{(n)}$ characterize 
the ``conventional'' nonlinear response. The scattering
of two photons off the same atom is for example 
determined by ${\cal P}^{(2)}$. 
It should be emphasized here, that cumulants containing operators of 
different atoms
vanish, since 
it was assumed that 
atom--atom correlations can be built up only by the quantized
radiation field. As a consequence each term ${\cal P}^{(n)}$ scales
only linearly with the number of atoms $N$. 
Thus ``conventional'' nonlinear interactions of increasing order require
increasing photon densities or large coupling constants $g_k$.

Franson {\it et al.} argued in \cite{Fr2} that a nonlinear phase shift
between two photons could emerge even if the ``conventional'' nonlinear
couplings, characterized by the higher-order cumulants in eq.(\ref{S}),
are negligible. Such phase shifts should arize 
from exchange interactions 
resulting from  to the symmetrization requirements imposed by the
bosonic nature of the photons. Let me therefore consider 
in the following the case were all higher-order cumulants are neglected.
In this situation ${\sf S}$ reduces to:
\begin{eqnarray}
{\sf S}\approx {\sf T}\,\exp
\Biggl\{\int\!\!{\rm d}1\!\int\!\!{\rm d}2\, 
{\hat a}_{k_1}^\dagger(\tau_1)\, {\cal P}(1,2)\, {\hat a}_{k_2}(\tau_2)
\Biggr\}.
\label{Sapp}
\end{eqnarray}
It should be emphasized that although the evolution operator (\ref{Sapp}) 
is bilinear in the photon operators, it takes fully into account any
exchange interaction. The implicit summation over mode indices accounts 
for processes where photon 1 is seen by atom $A$ and photon 2 by 
atom $B$ as well as the case where photon 1 is seen by atom $B$ and photon 2
by atom $A$.
It will now be shown that the conditional evolution $t_0\rightarrow t_1$
described by ${\sf S}$ cannot generate entanglement. I.e. any 
initially factorizable state will evolve into a factorizable state
after the interaction.


\section{State evolution}


In order to discuss the evolution of photons
described by ${\sf S}$ in (\ref{Sapp}), I consider the case of
the field initially being in a factorizable two-mode
state with at most one photon in each mode. $
|\phi(t_0)\rangle = |\phi_1\rangle \, |\phi_2\rangle\, 
 |\{0_k\}\rangle$ with 
\begin{eqnarray}
|\phi_1\rangle =\left(\alpha_1 +\beta_1 {\hat a}_{k_1}^\dagger\right)\, 
|0_1\rangle,\qquad
|\phi_2\rangle =\left(\alpha_2 +\beta_2 {\hat a}_{ k_2}^\dagger\right)\, 
|0_2\rangle.
\end{eqnarray}
Here $|0_1\rangle, |0_2\rangle$ are the vacuum states of modes
$k_1$ and $k_2$ and $|\{0_k\}\rangle$ is the vacuum state of all other 
modes. 

I proceed with discussing the evolution of the individual components of 
$|\phi(t_0)\rangle$. The vacuum component remains of course unaffected 
and it is 
sufficient to consider
\begin{eqnarray}
|\chi_1(t_1)\rangle &=& {\sf S}(t_1,t_0)\, |\chi_1(t_0)\rangle =
 {\sf S}(t_1,t_0)\,
{\hat a}_{k_1}^\dagger(t_0) 
|0\rangle,\label{s1}\\
&&{\rm and}\nonumber\\
|\chi_{1,2}(t_1)\rangle &=& {\sf S}(t_1,t_0)\, |\chi_{1,2}(t_0)\rangle
= {\sf S}(t_1,t_0)\, {\hat a}_{k_1}^\dagger(t_0)\, 
 a_{k_2}^\dagger(t_0) |0\rangle.\label{s2}
\end{eqnarray}
To formally calculate these expressions we make use of Wick's theorem,
which states that a time-ordered operator expression can be replaced by the
sum of all normally ordered expressions with all possible ``contractions''.
Contractions refer here to a replacement of any operator 
pairs ${\hat a}_{k'}^\dagger(\tau')$ and ${\hat a}_{k''}(\tau'')$ by the 
${\sf T}$-ordered propagator 
\begin{equation}
{\cal D}(1,2)=\Bigl\langle 0\Bigr|\, {\sf T}\,
{\hat a}_{k'}^\dagger(\tau_1) {\hat a}_{k''}(\tau_2)\, \Bigl|0\Bigr\rangle.
\end{equation}
We first note that since $t_0$ is the smallest time, the creation operators
${\hat a}_{k_1}^\dagger(t_0)$ and ${\hat a}_{k_2}^\dagger(t_0)$ 
in eqs.(\ref{s1}) and (\ref{s2}) can be included
in the ${\sf T}$-ordering. 
Since ${\sf S}\, {\hat a}_{k_1}^\dagger(t_0)$ and 
${\sf S}\, {\hat a}_{k_1}^\dagger(t_0)
{\hat a}_{k_2}^\dagger(t_0)$ respectively act on the vacuum state, out
of all normally ordered expressions only those
survive which have no photon annihilation operator left. 

Now  ${\sf S}\, {\hat a}_{k_1}^\dagger(t_0)$ 
can be expanded into a power series 
and Wick's theorem applied to each term.
This leads to the following
perturbation series
\begin{eqnarray}
|\chi_1(t_1)\rangle
&=&
\Biggl\{  {\hat  a}_{k_1}^\dagger(t_0)+ \int\!\!{\rm d}1\!\int\!\!{\rm d}2\, 
 {\cal D}(0,1)\Biggl[{\cal P}(1,2)
+\int\!\!{\rm d}3\!\int\!\!{\rm d}4\,
 {\cal P}(1,3){\cal D}(3,4){\cal P}(4,2)\nonumber\\
&&\quad
+
\int\!\!{\rm d}3\!\int\!\!{\rm d}4\!\!\int\!\!{\rm d}5\!\int\!\!{\rm d}6
\,  {\cal P}(1,3){\cal D}(3,4){\cal P}(4,5){\cal D}(5,6){\cal P}(6,2)
+\cdots\Biggr]\,{\hat a}_{k''}^\dagger(\tau_2)\,\Biggr\} |0\rangle\label{c1},
\end{eqnarray}
where $|0\rangle$ denotes the vacuum of all field modes.
The first term results from contractions of photon operators 
within ${\sf S}$. The other
terms arise 
from all possible contractions of ${\hat a}_{k_1}^\dagger$ with operators
from ${\sf S}$.

Eq.(\ref{c1}) can be given the compact form
\begin{eqnarray}
|\chi_1(t_1)\rangle
&=& \Biggl[{\hat a}_{k_1}^\dagger(t_0)
+
\int\!\!{\rm d}1\,\int\!\!{\rm d}2\, 
{\cal D}(0,1)\, \Pi(1,2)\, {\hat a}_{k''}^\dagger(\tau_2)\Biggr]|0\rangle
\end{eqnarray}
where $\Pi(1,2)$ is the solution to the linear integral equation
(Dyson equation)
\begin{eqnarray}
\Pi(1,2)={\cal P}(1,2) + \int\!\!{\rm d}2\!\int\!\!{\rm d}3\, 
{\cal P}(1,3)\, {\cal D}(3,4)\, \Pi(4,2).\label{dys}
\end{eqnarray}
In fact one easily verifies that an interactive solution of this equations
generates the whole perturbation series of (\ref{c1}).
That the quantum evolution can formally be solved 
in such a simple way is not surprising since the system is linear.
Eq.(\ref{dys}) describes nothing else than multiple scattering of the 
incoming photon at the atoms with all nonlinearities being absent.
In a diagrammatic language, the Dyson equation (\ref{dys})
corresponds to a sum of chain-like diagrams without branching
or merging. 

In a similar way as above one can proceed with
${\sf S}\, {\hat a}_{k_1}^\dagger {\hat a}_{k_2}^\dagger$.
In this case contractions only within ${\sf S}$
generate a term proportional to the product 
${\hat a}_{k_1}^\dagger {\hat a}_{k_2}^\dagger$
similar to the first term in eq.(\ref{c1}). Then two series of terms
emerge where either ${\hat a}_{k_1}^\dagger$ 
or ${\hat a}_{k_2}^\dagger$ is contracted with operators from ${\sf S}$. 
These leads
to expressions identical to the higher-order terms in (\ref{c1}) 
multiplied with
either ${\hat a}_{k_1}^\dagger$ or ${\hat a}_{k_2}^\dagger$. Finally 
there is a series
of terms resulting of contractions of {\it both} ${\hat a}_{k_1}^\dagger$ and 
${\hat a}_{k_2}^\dagger$ with operators from ${\sf S}$. 
This yields
\begin{eqnarray}
&&|\chi_{12}(t_1)\rangle
=
\Biggl\{   {\hat a}_{k_1}^\dagger(t_0)
 {\hat a}_{k_2}^\dagger(t_0)+\nonumber\\
&&
+ \int\!\!{\rm d}1\!\int\!\!{\rm d}2\, 
 {\cal D}(0',1)\biggl[{\cal P}(1,2)
+\int\!\!{\rm d}3\!\int\!\!{\rm d}4\,
 {\cal P}(1,3){\cal D}(3,4){\cal P}(4,2)
+\cdots\biggr]\, {\hat a}_{k_2}^\dagger(t_0)\,
{\hat a}_{k''}^\dagger(\tau_2)+\nonumber\\
&&
+ \int\!\!{\rm d}1\!\int\!\!{\rm d}2\, 
 {\cal D}(0'',1)\biggl[{\cal P}(1,2)
+\int\!\!{\rm d}3\!\int\!\!{\rm d}4\,
 {\cal P}(1,3){\cal D}(3,4){\cal P}(4,2)
+\cdots\biggr]\, {\hat a}_{k_1}^\dagger(t_0)\,
{\hat a}_{k''}^\dagger(\tau_2)+\\
&&
+\enspace \int\!\!{\rm d}1\!\int\!\!{\rm d}2\, 
 {\cal D}(0',1)\biggl[{\cal P}(1,2)
+\int\!\!{\rm d}3\!\int\!\!{\rm d}4\,
 {\cal P}(1,3){\cal D}(3,4){\cal P}(4,2)
+\cdots\biggr]\,\times\nonumber\\
&&\enspace\times 
\int\!\!{\rm d}{\tilde 1}\!\int\!\!{\rm d}{\tilde 2}\, 
 {\cal D}(0'',{\tilde 1})\biggl[{\cal P}({\tilde 1},{\tilde 2})
+\int\!\!{\rm d}{\tilde 3}\!\int\!\!{\rm d}{\tilde 4}\,
 {\cal P}({\tilde 1},{\tilde 3}){\cal D}({\tilde 3},{\tilde 4})
{\cal P}({\tilde 4},{\tilde 2})
+\cdots\biggr]\,{\hat a}_{k''}^\dagger(\tau_2)\,
{\hat a}_{{\tilde k}''}^\dagger({\tilde \tau}_2)
\,\Biggr\} |0\rangle.\nonumber
\end{eqnarray}
Here $0'$ and $0''$ stand for $\{t_0,k_1\}$ and $\{t_0,k_2\}$ respectively.
This expression can again be brought into a compact form
\begin{eqnarray}
&&|\chi_{12}(t_1)\rangle
= {\hat a}_{k_1}^\dagger(t_0)
 {\hat a}_{k_2}^\dagger(t_0)\, |0\rangle\nonumber\\
&&+\int\!\!{\rm d}1\! \int\!\!{\rm d}2\, 
{\cal D}(0',1)\, \Pi(1,2) {\hat a}_{k''}^\dagger(\tau_2)
 {\hat a}_{k_2}^\dagger(t_0)
|0\rangle+
\int\!\!{\rm d}1\! \int\!\!{\rm d}2\, 
{\cal D}(0'',1)\, \Pi(1,2) {\hat a}_{k''}^\dagger(\tau_2)
 {\hat a}_{k_1}^\dagger(t_0)
|0\rangle\nonumber\\
&&+\int\!\!{\rm d}1\! \int\!\!{\rm d}2\, 
{\cal D}(0',1)\, \Pi(1,2)
\int\!\!{\rm d}{\tilde 1}\! \int\!\!{\rm d}{\tilde 2}\, 
{\cal D}(0'',{\tilde 1})\, \Pi({\tilde 1},{\tilde 2})\,
\, {\hat a}_{k''}^\dagger(\tau_2)\, {\hat a}_{{\tilde k}''}^\dagger
({\tilde \tau}_2)\, |0\rangle.
\nonumber
\end{eqnarray}
One immediately recognizes that $|\chi_{12}(t_1)\rangle$ can be written as
\begin{eqnarray}
|\chi_{12}(t_1)\rangle &=&\quad
\Biggl[{\hat a}_{k_1}^\dagger(t_0)
+
\int\!\!{\rm d}1\int\!\!{\rm d}2\, 
{\cal D}(0',1)\, \Pi(1,2)\, {\hat a}_{k''}^\dagger(\tau_2)\Biggr]\nonumber\\
&&\otimes\Biggl[{\hat a}_{k_2}^\dagger(t_0)
+
\int\!\!{\rm d}{\tilde 1}\int\!\!{\rm d}{\tilde 2}\, 
{\cal D}(0'',{\tilde 1})\, \Pi({\tilde 1},{\tilde 2})\, 
{\hat a}_{{\tilde k}''}^\dagger({\tilde \tau}_2)\Biggr]\,  |0\rangle
\end{eqnarray}
The  evolution of  
$|\phi\rangle$ from $t_0$ to $t_1$ is hence given by
\begin{eqnarray}
&&\qquad|\phi(t_0)\rangle = \left(\alpha_1 +\beta_1 {\hat a}_{k_1}^\dagger
\right)\, \left(\alpha_2 +\beta_2 {\hat a}_{k_2}^\dagger\right)\,
|0\rangle\nonumber\\
&&\qquad\qquad\qquad\downarrow\nonumber\\
|\phi(t_1)\rangle =&&\left[\alpha_1 
+\beta_1\Biggl({\hat a}_{k_1}^\dagger(t_0)
+
\int\!\!{\rm d}1\int\!\!{\rm d}2\, 
{\cal D}(0',1)\, \Pi(1,2)\, {\hat a}_{k''}^\dagger(\tau_2)\Biggr)
\right]\otimes\\
&&\left[\alpha_2 +\beta_2\Biggl({\hat a}_{k_2}^\dagger(t_0)
+
\int\!\!{\rm d}{\tilde 1}\int\!\!{\rm d}{\tilde 2}\, 
{\cal D}(0'',{\tilde 1})\, \Pi({\tilde 1},{\tilde 2})\, 
{\hat a}_{{\tilde k}''}^\dagger({\tilde \tau}_2)\Biggr)
\right]\, \Bigl|0\Bigr\rangle.\nonumber
\end{eqnarray}
Thus if the process starts with a factorizable state with 
photons in distinguishable modes, i.e. if
$(\alpha_1 +\beta_1 {\hat a}_{k_1})|0\rangle$ is orthogonal to
$(\alpha_2 +\beta_2 {\hat a}_{k_2})|0\rangle$ and if the process
generates photons in distinguishable modes, i.e. if
\begin{eqnarray}
\left[\alpha_1 +\beta_1\Bigl({\hat a}_{k_1}^\dagger
+
\int\!\!\int\, 
{\cal D}\, \Pi\, 
{\hat a}_{{k}''}^\dagger\Bigr)
\right]\, \Bigl|0\Bigr\rangle\qquad{\rm and}\qquad
\left[\alpha_2 +\beta_2\Bigl({\hat a}_{k_2}^\dagger
+
\int\!\!\int\, 
{\cal D}\, \Pi\, 
{\hat a}_{{\tilde k}''}^\dagger\Bigr)
\right]\, \Bigl|0\Bigr\rangle\nonumber
\end{eqnarray}
are orthogonal,
 then 
the generated state vector remains factorizable. 


\section{Conclusion}


In the present note I have shown it is not possible to generate 
entanglement between photons using {\it solely} exchange interactions in a 
large ensemble of atoms, if the atoms are left in the same quantum
state after the interaction as they were initially. From a diagrammatic
point of view entanglement between photons can not be generated if
all possible diagrams are chain-like. To produce entanglement
non-trivially connected diagrams are needed, as emerge for example 
from nonlinear atomic responses or from atom--atom interactions due to
e.g. dipole-dipole or collisional interactions.


\section*{acknowledgment}


I would like to thank Mikhail Lukin and Tomas Opatrny for stimulating
discussions on the subject and the Institute for
Atomic and Molecular Physics for the hospitality.
The financial support of the 
Deutsche Forschungsgemeinschaft is highly appreciated.


\end{document}